\documentclass{elsart}
\usepackage{epsfig}
\begin{document}

\begin{frontmatter}



\title{Signals of bimodality in the fragmentation of Au quasi-projectiles}


\author [bo]{M. Bruno}, \author [lpc]{F. Gulminelli\thanksref{iuf}},
\author [bo]{F. Cannata}, \author[bo]{M. D'Agostino\corauthref{cor}}, 
\author[lnl]{F. Gramegna},
\author[bo]{G. Vannini}

\address[bo]{Dipartimento di Fisica dell'Universit\`{a} and INFN,
Bologna, Italy}
\address[lpc]{LPC Caen IN2P3-CNRS/EnsiCaen et Universit\`{e}, Caen, France}
\address[lnl] {INFN, Laboratori Nazionali di Legnaro, Italy}
\thanks[iuf]{Member of the Institut Universitaire de France}
\corauth[cor]{Corresponding author.\\
\hfill {\it  E-mail address: } dagostino@bo.infn.it (M. D'Agostino)}

\begin{abstract}
Signals of bimodality have been investigated in experimental data of
quasi-projectile decay produced in Au+Au collisions at 35 AMeV. This
same data set was already shown to provide several signals
characteristic of a first order, liquid-gas-like phase transition.
Different event sortings proposed in the recent literature are
analyzed. A sudden change in the fragmentation pattern is revealed
by the distribution of the charge of the largest fragment,
compatible with a bimodal behavior.
\end{abstract}

\begin{keyword}
NUCLEAR REACTIONS Au(Au,X), E = 35 AMeV \sep Measured fragment
charge distributions \sep Excited quasi-projectile decay \sep
Reweighting procedure \sep Deduced bimodal behavior of largest
fragment charge

\PACS{
      {05.70.Fh}{Phase transitions: general studies} \sep
      {25.70.-z}{Low and intermediate energy heavy-ion reactions}
     } 
\end{keyword}
\end{frontmatter}
\section{Introduction}
The existence of different phases for infinite nuclear matter is
predicted by theoretical calculations since the early
80's~\cite{history,SMM,gross,rass}. Then, the possibility of
observing a nuclear liquid-gas phase transition in the laboratory
has been deduced from several experimental observations associated
to the multi-fragmentation of finite nuclei. These observations
indicate the occurrence of a change of state  in finite nuclei,
which is interpreted to be the finite system counterpart of a phase
transition~\cite{chomaz}.

Many of these signals are qualitative, and therefore can\-not give
information on the detailed path (in terms of pressure, volume,
temperature, isospin) followed by the system from one phase to the
other. Other signals give in principle quantitative information, but
can be distorted. Indeed the products of the reactions are detected
asymptotically and not at the production time, and therefore they
need to be corrected for secondary decay. These corrections are, at
least partially, model dependent and induce systematic errors which
are difficult to estimate quantitatively~\cite{rely}.

To overcome these difficulties, it is important to perform a
systematic study of different phase transition signals. The best
would be to exploit new generation 4$\pi$ apparatuses, in order to
be able to investigate simultaneously several signals at the same
time, with the same experimental data samples, and with a complete
or quasi-complete detection. Waiting for these new apparatuses, some
of the signals indicating a phase transition have been obtained with
measurements performed by the Multics~+~Miniball
multi-detectors~\cite{det}. In the last few years we have
investigated in detail the properties of quasi-projectiles detected
in Au + Au reactions at 35 AMeV, with a fixed source charge, and at
different excitation energies~\cite{praga}. The following signals
have been obtained:
\begin {enumerate}
\item the average size of the heaviest fragment (tentatively associated to the
order parameter) decreases for increasing excitation energy of the
nuclear system~\cite{dag} with a power law distribution of exponent
$\beta\approx 0.31$;
\item temperature measurements result compatible~\cite{rely,mil} with a "plateau" in
the caloric curve~\cite{caloric}; \item critical exponents have been
extracted~\cite{dag}, close to the values expected within the
liquid-gas universality class;
\item the size distribution presents a scaling
"\`a la Fisher"~\cite{critical};
\item interaction energy fluctuations, corrected for si\-de-fee\-ding,
were shown to overcome the statistical expectation in the canonical
ensemble, corresponding to a negative branch of the microcanonical
heat capacity for a system in thermodynamical
equilibrium~\cite{negative,negth}.
\end {enumerate}

Several of these signals are consistent with the findings of other
experimental collaborations with different data sets \cite{ma,coll}.
In particular, the last two signals have been confirmed in central
reaction measurements ~\cite{rely,dag2,bruno,leneindre}. Some of
these behaviors were also observed in other finite physical systems
undergoing a transformation interpreted as a first order phase
transition, namely in the melting of atomic clusters~\cite{negclu}
and in the fragmentation of hydrogen clusters~\cite{neg3}.

Recently~\cite{PRE}, a new topological observable has been proposed
to recognize first order phase transitions. When a finite system
undergoes such a transition, the most probable value of the order
parameter changes discontinuously, while the associated distribution
is bimodal close to the transition point, i.e. it shows two separate
peaks, corresponding to the two different
phases~\cite{binder,dasgupta}. In the case of transitions with a
finite latent heat, this behavior is in agreement with the Yang-Lee
theorem for the distribution of zeroes of the canonical partition
sum in the complex temperature plane~\cite{zeroes}, and equivalent
to the presence of a curvature anomaly in the microcanonical entropy
$S(E)$~\cite{gross,intruder}.

Since many different correlated observables can act as order
parameters in a finite system, the task is to choose an order
parameter which can be accessible to experiments~\cite{WCI}. This is
the case for observables related to the measured charges. The Indra
collaboration~\cite{indra} has proposed as order parameter the
 variable Z$_{sym}=\frac {Z_1-Z_2} {Z_1+Z_2}$, where
Z$_1$ and Z$_2$ are the charge of the largest and the second largest
fragments detected, in each event, in the decay of an excited
source. An indication of a bimodal distribution was obtained for
this quantity. Signals of bimodality in different observables have
been obtained in experiments with different projectile-target
combinations, and in different energy ran\-ges~\cite{ma,WCI,bell}.
Also experimental results of Aladin group~\cite{traut} show a
bimodal distribution of the 3-fragments difference ($Z_1 - Z_2
-Z_3$)~=~$\Delta$Z.

In Refs.~\cite{WCI,traut} it has been pointed out that the variables
Z$_{sym}$ and $\Delta$Z can present a spurious bimodality in small
three-dimensional percolation lattices close to the percolation
threshold. This behavior is due to finite size, and makes the
bimodality in asymmetry variables an ambiguous signature of the
transition.  On the other side, the size $A_1$ or charge $Z_1$ of
the largest fragment have distributions which for any lattice size
are consistent with the critical percolation
behavior~\cite{campi,big}. These observables were then suggested as
more apt to discriminate between a first order phase transition, a
critical phenomenon, and a smooth cross-over, even if some
ambiguities in the interpretation of this signal still
exist~\cite{lop,aich}.

 In this paper we investigate whether signals
of bimodality for the charge of the largest fragment can emerge from
our data.

\section{The experiment} \label{exp}

The measurements and the analysis have been extensively described
elsewhere~\cite{dag}. Here we recall that the measurements were
performed at the K1200-NSCL Cyclotron of the Michigan State
University. The Multics and Miniball arrays~\cite{det} were coupled
to measure light charged particles (Z$\le$~2) and fragments
(Z$\ge$~3) with a geometrical acceptance of the order of 87\% of
4$\pi$. The events have been recorded if at least two different
modules have been fired. The selection of the quasi-projectile (QP),
 made in
Refs.~\cite{rely,praga,dag,critical,negative,dag2,bruno}, required
the velocity of the largest fragment in each event to be at least
75\% of the beam velocity. After a shape analysis~\cite{cugn},
events with a total forward detected charge larger than 70\% of the
Au charge were considered. The complete source was obtained by
doubling the forward emitted light particles, in order to minimize
the contamination of particles emitted by a possible mid-velocity
source. At the end of this procedure, only events with total charge
within 10\% of the Au charge were considered for the analysis, in
order to study the decay of a well detected constant size source, in
a wide range of excitation energies.

\begin{figure}[htbp]
\begin{center}
{\includegraphics[height=8cm]{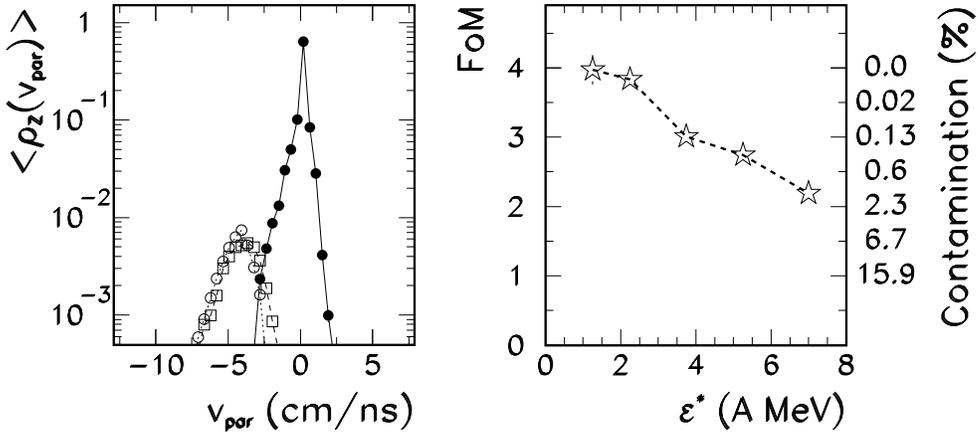}} 
\caption{
Left panel:
Charge density distribution of QP Au+Au events, as a function of the
fragment velocity, along an axis parallel to the QP velocity. The
continuous line (full points) represents the charge density for
fragments accepted for the QP, the dashed line (open squares) for
fragments rejected, the dotted line (open circles) is a filtered
simulation of a QT source symmetric to the experimentally detected
QP.  Right panel: Factor of Merit (FoM) and maximal contamination of
QT into QP source as a function of the excitation energy per nucleon
$\varepsilon$*. The dashed line is drawn to guide the eye. For more
details see text. When not visible, error bars are smaller than the
size of the points.} \label{vpar}\end{center}
\end{figure}

In order to visualize the source characteristics in the selected
events, the fragment (Z$\geq 3$) charge density
distribution~\cite{lecolley} is shown in Fig.~\ref{vpar} as a
function of the fragment velocity in the QP reference frame. The
ensemble averaged charge density $\langle\rho_Z(v_{par})\rangle$ is
defined as

$$ \langle\rho_Z(v_{par})\rangle = \Big\langle {\sum Z(v_{par}) \over
\sum Z} \Big\rangle $$

where $Z(v_{par})$ is the event-by-event distribution in the
velocity $v_{par}$ for the fraction of collected charge and the sum
extends over all the fragments. This observable represents the
distribution of the collected charge of fragments along the
direction of the QP velocity. In  the left panel of
Fig.~\ref{vpar} the continuous line (full points) represents the
charge density for fragments accepted for the QP, the dashed line
(open squares) for fragments backward emitted in the ellipsoid
reference frame. The open circles show a filtered simulation of a QT
source symmetric to the QP. The consistency between these two latter
curves demonstrates that a purely binary dynamics exhausts, in first
approximation, the totality of the emitted fragment charge for the
set of selected events. QP and QT can be easily recognized, showing
that the imposed conditions are effective in selecting events where
the contamination of a mid-velocity source is negligible (for more
details see Refs.~\cite{rely,dag}). It also appears clearly from
the left panel of Fig.~\ref{vpar} that the velocity distance
between the two sources is large enough to insure a negligeable
contamination of QT decays in the reconstructed QP source.

To have a quantitative insight on this contamination we have
considered the charge density distributions in different bins of
excitation energy, as in Fig.~1 of ~\cite{dag}. For these
distributions we have calculated a quantity similar to the one used
to give a measurement of the discrimination in particle
identification~\cite{win}, the Factor of Merit (FoM). The FoM used
here is $ FoM = d_{QP-QT} / (\sigma_{QP}+\sigma_{QT})$, i.e. the
ratio between the distance of the QP and QT average velocities over
the sum of their standard deviations (assuming a gaussian behavior).
In the right panel of Fig.~\ref{vpar} the FoM is shown, together
with the evaluated contaminations of QT into the QP source (right
scale). The contamination has been calculated in the hypothesis of
two distributions of equal height and standard deviation, as it
would be the case of an ideal detection of QP and QT for a symmetric
reaction. It appears well below a few percent.

Since, however, the apparatus efficiency is lower for QT with
respect to QP heavy products the height of the distribution of the
QT is much lower than the one of the QP, as it is clear from the
left panel of Fig.~\ref{vpar}. If one therefore calculates the
contamination with the real QP and QT distributions in all the bins
of excitation energy one obtains values well below the values shown
in Fig.~\ref{vpar}. The maximal contamination, obtained for the
highest excitation energy, results of the order of 0.6\%.

\begin{figure}[htbp]
\begin{center} \resizebox{.8\columnwidth}{!}{\includegraphics{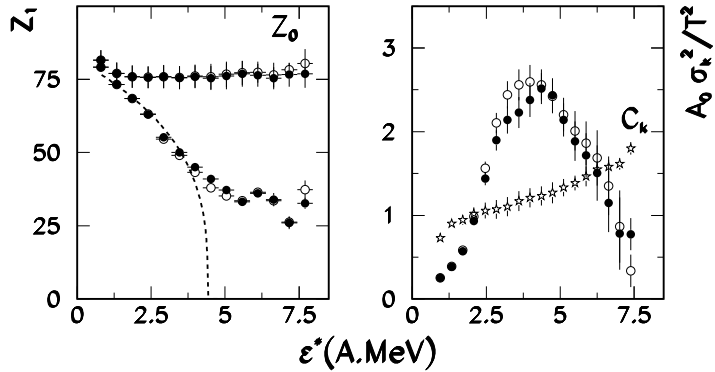}}
\caption{Left panel: charge of the largest fragment Z$_1$ as a
function of the QP calorimetric excitation energy. The symbols refer
to event selected with (full points) and without (open circles) the
constraint on the velocity of the largest fragment. The dashed line
is a power law with exponent $\beta$ = 0.31. The total charge of the
source, as a function of the excitation energy, is also shown for
the two analyses. The right panel shows the normalized partial
energy fluctuations for QP events selected with (full points) and
without (open points) the constraint on the velocity of the largest
fragment. The estimation for the canonical heat capacity C$_k$ is
also shown~\protect\cite{negative}.} \label{cneg}\end{center}
\end{figure}

The characteristics of the QP events have been examined by analyzing
the isotropy of the fragment angular distribution  in the
quasi-projectile reference frame~\cite{negative,dag2}, and by
comparing the data to predictions of a statistical
multifragmentation model~\cite{SMM}. The general conclusion is that
an important degree of equilibration appears to be reached by the
excited quasi-projectile sources in the whole range of excitation
energies. For more details, see
Refs.~\cite{rely,praga,dag,negative,dag2}.

One of the conditions used to characterize the QP, i.e. that the
velocity of the heaviest fragment is larger than 75\% of the
projectile velocity, in the analysis here presented has been
released, in order to minimize the correlation to the variables we
want to study and to allow a better comparison to the results of
Ref.~\cite{indra}, where this condition was not imposed. In this
paper we have also restricted the condition on the total forward
detected charge to 80\% of the Au charge, again for consistency with
the choice taken in Ref.~\cite{indra}. These modifications do not
affect the distribution shown in Fig.~\ref{vpar} nor significally
change the signals of phase transition. To quantify this statement,
we present in Fig.~\ref{cneg} the power-law behavior of the average
charge of the largest fragment, as a function of the excitation
energy, and the normalized partial-energy fluctuations, leading to
the estimate of a negative branch for the microcanonical heat
capacity~\cite{negative}. The power law in the Z$_1$ distribution
and the partial energy fluctuations are very little affected by the
different selection conditions.

As a final remark we should stress that only a fraction of well
detected peripheral collisions can be interpreted as the independent
statistical decay of two isotropic sources. For instance within the
Indra apparatus it has been pointed out that for 80 AMeV Au+Au
collisions, these events represent about the 20\% of the total
number of complete events~\cite{leneindre}, and this number depends
on the selection criteria adopted~\cite{bonnet}. In our case the
statistical events represent about 30\% of the measured
events~\cite{dag}. The difference in the percentage of statistical
events could also be due to the different trigger conditions of four
(respectively: two) modules fired, used in Indra and in our
measurements.

\section{Signals of bimodality}

In the liquid-gas phase transition, the largest fragment detected in
each event is a natural order parameter, because of its correlation
with the particle density in the grancanonical
ensemble~\cite{PRE,big}. The asymmetry variable Z$_{sym}$ proposed
in Ref.~\cite{indra,rivet} in turn is correlated to the largest
cluster size, and should bring further information on the global
fragmentation pattern. This means that it should be possible to
observe a bimodal distribution for the charge of the largest
fragment or the asymmetry, if one considers a system close to the
transition temperature.

To perform a meaningful analysis of a possible bimodal pattern in the
Z$_{sym}$ observable, one needs to properly treat fission events.
Indeed for a quasi-gold nucleus the binary fission, recognized for $Z_1$ and
$Z_2$ values around 40 charge units, would lead to small values of
Z$_{sym}$, as multifragmentation events.

Fission fragments can be recognized by looking at the product
of their charges, exceeding 900, and they can be either rejected
as in our analyses (see Ref.\cite{dag}) or re-clusterized (as
in Ref.\cite{indra}).
The main effect of both choices is to eliminate a spurious peak at 
(Z$_{sym} \approx 0$, $Z_1\approx 40$).

\begin{figure}[htbp]
\begin{center}
\resizebox{0.7\columnwidth}{!}{\includegraphics{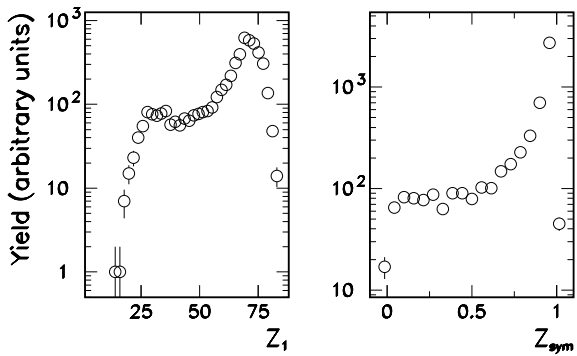}}
\caption{Distribution of the largest fragment charge Z$_1$, (left
panel) and of the asymmetry between the two largest fragments
Z$_{sym}$ (right panel) for the whole set of QP events. When not
visible, error bars are smaller than the size of the points.}
\label{whole}
\end{center}
\end{figure}

The global distributions of Z$_1$ and Z$_{sym}$ are shown in
Fig.~\ref{whole}, for all QP events selected as explained in section
2. Because of the impact parameter geometry, this distribution is
clearly dominated by peripheral collisions at low deposited energy,
leading to a heavy Z$_1\approx 75$ residue with a large asymmetry
Z$_{sym}\approx 0.9$.

We can however also notice the presence of another bump,
corresponding to much lighter residues and much more symmetric
fragmentation patterns. For this bump to be interpreted as an
indication of bimodality, we should show that:
\begin{itemize}
\item the two different decay patterns
can be obtained in the de-excitation of the same source,
\item they correspond to the same temperature.
\end{itemize}
These points are discussed in the next subsections.

\subsection {Size of the QP source}

Let us first concentrate on the first point about the source
definition. As we have already stressed in the last section and
shown in Fig.~\ref{cneg}, we are considering only events with a
detected charge in the forward QP hemisphere close to the original
$Au$ charge. This guarantees a good detection, but does not
\begin{figure}[h]
\begin{center}
\resizebox{0.65\columnwidth}{!}{\includegraphics{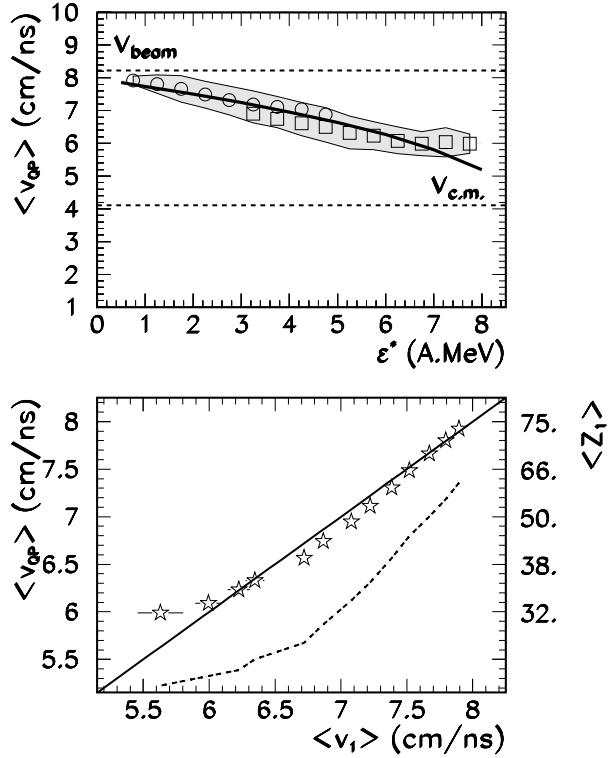}}
\caption{Upper panel: Average velocity of the QP source, in the
laboratory reference frame, as a function of the excitation energy
(grey contour). Open circles and squares refer to Z$_1 >$ 50 and $<$
50, respectively. The continuous line is obtained by energy and
momentum conservation in a two body QP and QT kinematics~\cite{dag}.
Beam and center of mass velocities are indicated by the dashed
lines. Lower panel: Average QP velocity as a function of the average
velocity of the largest fragment at the same $\varepsilon$*. The
dashed line is calculated in a neck-like hypothesis (see text). The
values of the average charge of the largest fragment
$\langle$Z$_1\rangle$ are shown in the right scale.} \label{twobody}
\end{center}
\end{figure}
constrain the reaction mechanism or the number of sources, since the
system is symmetric. Fig.~\ref{vpar} shows that the selected events
are consistent with a purely binary kinematics. This in principle
already implies that the bump at low charge shown by
Fig.~\ref{whole} cannot be ascribed to a reduced size of the excited
source. Furthermore Fig.~\ref{vpar} and the calculated values in
$\S$~\ref{exp} show that the the contamination of the QT source is
everywhere negligible, indicating that the bump at low charge in
Fig.~\ref{whole}, corresponding to the most dissipative events,
cannot be responsible of a decrease of the QP source size. To check
this point, in Fig.~\ref{twobody} we plot the velocity in the
laboratory frame of the QP source as a function of the excitation
energy (grey contour). Open circles and squares correspond to cuts
of Z$_1>$ 50 and Z$_1 <$ 50, that is to the two bumps in the left
panel of Fig.~\ref{whole}. The source velocity expected for a two
body (QP-QT) kinematics, obtained via energy and momentum
conservation in the hypothesis of an equal sharing of the excitation
energy by the two equal mass collision partners, is indicated by the
thick line~\cite{dag}. We can see that both decay modes correspond
to a source velocity compatible with a binary mechanism. In
particular in the $\varepsilon$* region where the two modes are
simultaneously present, their velocities are close, meaning that
they can be attributed to a same source.

A further check is given in the lower part of Fig.~\ref{twobody},
which displays the reconstructed average QP velocity as a function
of the measured Z$_1$ velocity. In this figure the dashed line gives
the result of a simulation where all the forward collected fragments
except Z$_1$ are supposed to be emitted from an isotropic neck-like
source located at the center of mass, with the constraints of energy
and momentum conservation. We can see that even for the lowest
velocities, corresponding to the highest kinetic energy loss, the
measured correlation is very close to the diagonal (full line)
corresponding to a perfect emission isotropy respect to Z$_1$, while
the measured kinematics is never compatible with the emission from a
neck. Therefore the lower part of Fig.~\ref{twobody} shows that the
more fragmented decay pattern 
is not due to the presence in the data
set of a different reaction mechanism dominated by matter stopped in
the center of mass.

As a conclusion we can safely consider the data of Fig.~\ref{whole}
as characteristic of the de-excitation of constant size source in a
wide range of excitation energies.

\subsection {Data sorting}

Let us now come to the second central question of data sorting. The
global distributions of Fig.~\ref{whole} reflect the excitation
energy deposit imposed by the dynamics of the entrance channel, and
cannot be considered as belonging to a single statistical ensemble.
If a sorting cannot be avoided, it is also clear that the shape of
the distributions will depend on the sorting choice.

The two de-excitation modes visible in Fig.~\ref{whole} are
associated to different excitation energies (see
Fig.~\ref{twobody}). If they represent two different phases, this
means that the associated phase transition should have a non zero
latent heat, as is the case for standard liquid-gas. Therefore, the
sorting variable should not impose a too strong constraint on the
deposited energy, such that the two phases can be accessed in the
same bin. In particular, in the liquid-gas phase transition, Z$_1$
is known to be bimodal in the canonical ensemble which allows huge
energy fluctuations, while no bimodality is observed in the
microcanonical ensemble with fixed energy~\cite{PRE}.

To search for a possible bimodal behavior, we should then in
principle sort the data in temperature bins, i.e. in canonical
ensembles. This is not possible experimentally, but we can choose a
sorting variable allowing for relatively large energy fluctuations,
as it is needed to explore two phases separated by a non zero latent
heat. Moreover, as suggested by previous papers~\cite{indra,rivet},
the sorting observable should better not be auto-correlated with
fragments and light particles emitted by the QP source. To fulfill
these requirements, in~\cite{indra}, the transverse energy
E$_{12}^t$ = $\sum_i E_i sin^2 (\theta_i)$ of the light particles
($Z_i\leq2$) emitted by the quasi-target source has been proposed,
such particles being very efficiently detected even in the backward
direction. The QT had been much larger than the QP, this sorting
could be considered as a canonical one. In the present case,
however, this sorting can be rather assimilated to an impact
parameter sorting.

In our data the transverse energy extends up to $\approx$~400 MeV,
but the statistics for the higher values is poor. Therefore, after
having divided the total range in nine equally spaced bins (width 45
MeV) we have considered for the analysis only the first six bins
which have sufficient statistics.

\begin{figure}[htbp]
\begin{center}
\resizebox{0.7\columnwidth}{!}{\includegraphics{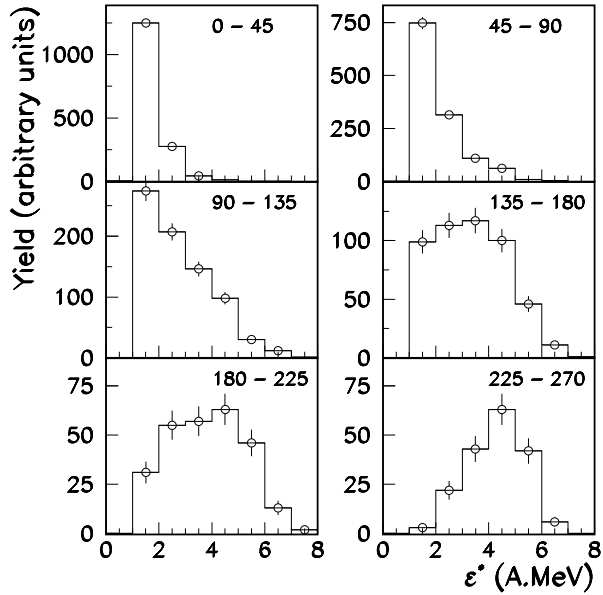}}
\caption{Calorimetric $\varepsilon$* distribution in the six bins of
transverse energy E$_{12}^t$, indicated in the panels as intervals.}
\label{estar}
\end{center}
\end{figure}

The excitation energy constraint implied by this sorting is explored
in Fig.~\ref{estar}, which shows the distribution of the
calorimetric $\varepsilon$* in the six transverse energy bins. We
can notice from this figure that the variables E$_{12}^t$ and
\begin{figure}[htbp]
\begin{center}
\resizebox{0.7\columnwidth}{!}{\includegraphics{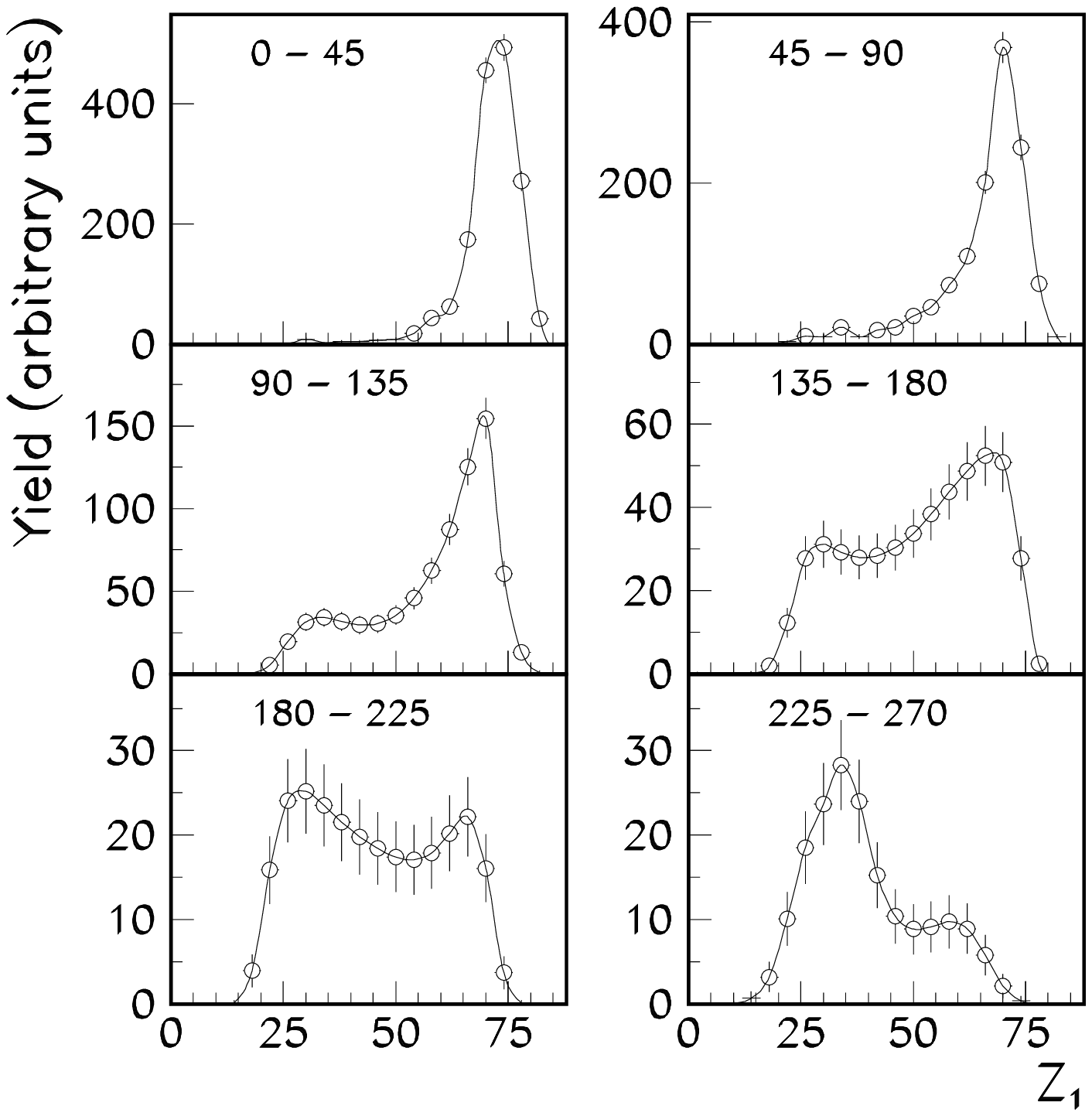}}
\caption{Distribution of the Z$_1$ variable in the different
transverse energy bins. The intervals of the transverse energy are
indicated in the panels. When not visible, error bars are smaller
than the size of the points.} \label{projbig}
\end{center}
\end{figure}
\begin{figure}[htbp]
\begin{center}
\resizebox{0.45\columnwidth}{!}{\includegraphics{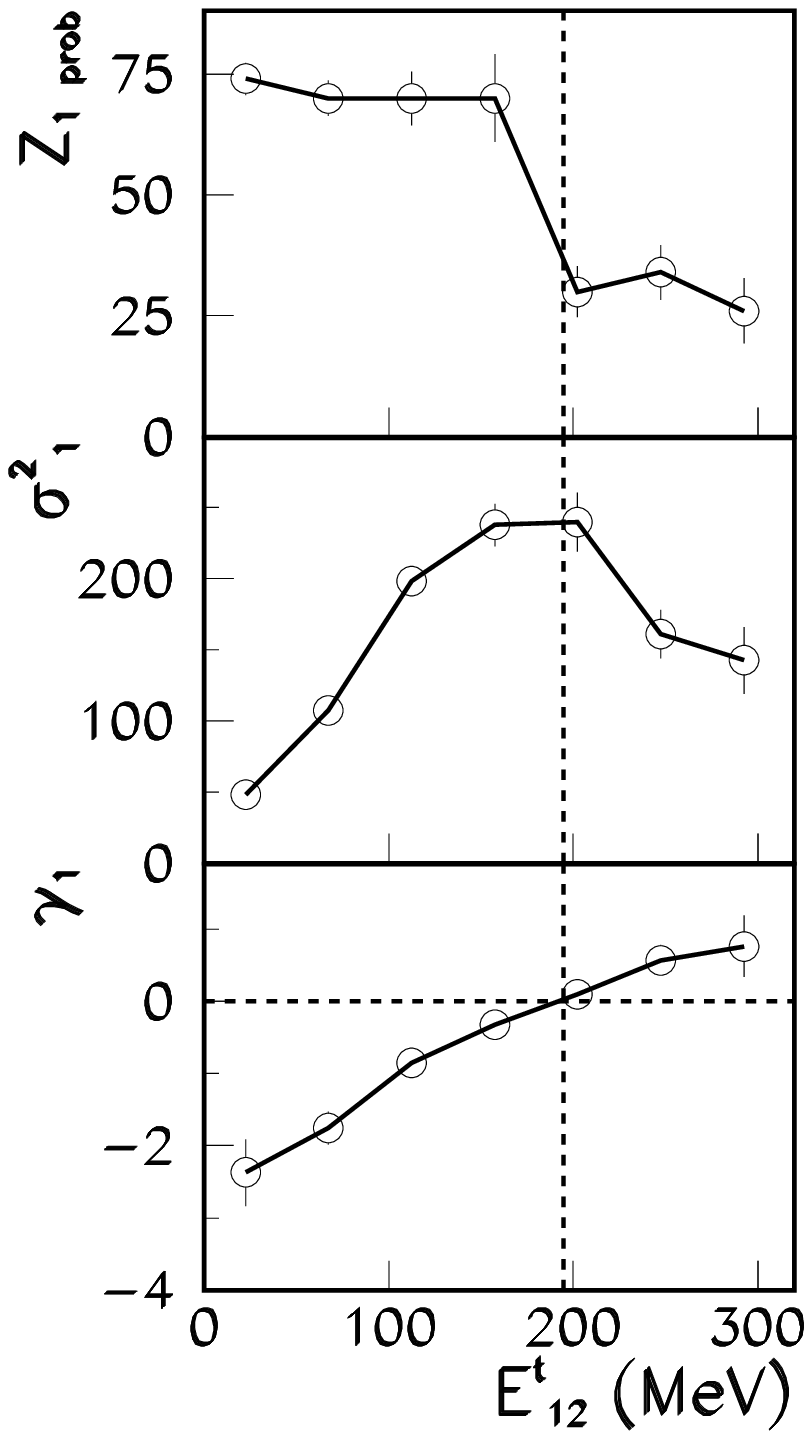}}
\caption{Evolution with E$_{12}^t$ of the most probable value of
Z$_1$ (upper panel), the variance $\sigma_1^2$ (middle panel) and
the skewness $\gamma_1$ of the distribution (lower panel). The full
lines are drawn to guide the eye. In this figure also the seventh
bin of the transverse energy is shown, despite of the low
statistics. When not visible, error bars are smaller than the size
of the points.} \label{jump}
\end{center}
\end{figure}
$\varepsilon$* are loosely correlated, and a relatively wide
distribution of $\varepsilon$* is obtained in most of the bins of
transverse energy. It is well known~\cite{rely} that the
calorimetric measurement is not perfect, and the incomplete
detection induces an uncertainty of the order 10\% on
$\varepsilon$*. However from Fig.~\ref{estar} it is clear that this
uncertainty does not affect the width of the distributions. The
sorting in E$_{12}^t$ bins cannot be therefore considered as a
microcanonical selection, where no bimodality would be expected a
priori.

The charge of the heaviest fragment Z$_1$ is plotted in
Fig.~\ref{projbig} for the six different transverse energy bins. We
can note some indication of bimodality since the largest fragment
size, peaked around $Z_1\approx$ 70 up to the fourth bin, shows a
maximum around $Z_1\approx 30$ in the sixth bin, passing through a
configuration (fifth bin) where a minimum in the probability appears
to be associated to the intermediate patterns, even if admittedly
the statistics should be improved. Such a behavior agrees with
previous findings for peripheral Xe + Sn and Au + Au
collisions~\cite{indra}, and with the expectations from a phase
transition. We have already mentioned that the considered sorting
corresponds to an impact parameter selection. Fig.~\ref{projbig}
then shows that the two different decay patterns are associated to
the same initial condition for the collision~\cite{aich}.

In Fig.~\ref{jump} the most probable value of Z$_1$ is shown,
together with the variance and the skewness $\gamma_1$ of Z$_1$
distribution, as a function of the transverse energy. The skewness
is defined as $\gamma_1 ={ \mu_3 / \sigma_1^3}$ where $\mu_3$ is the
third moment about the mean and $\sigma_1$ is the standard
deviation. The jump in the most probable value of Z$_1$ corresponds
to a maximum in the variance and a change of sign in the skewness
(dashed line in Fig.~\ref{jump}). The detailed shape of the
distributions shown in Figs.~\ref{estar} and \ref{projbig} obviously
depends on the (largely arbitrary) width of the transverse energy
bins, as well as on the choice of the sorting variable. The sudden
change of the skewness of the distribution, passing through a
configuration of maximal fluctuations, is however independent of
these choices.

\subsection {Towards a canonical sorting}

The use of a transverse energy sorting has allowed us to directly
compare to the previously analyses presented in the
literature~\cite{indra}. However, as already discussed, such a
sorting cannot be interpreted as a canonical sorting. Indeed the
absence of an explicit microcanonical constraint as shown in
Fig.~\ref{estar} does not guarantee that the energy fluctuations are
large enough to allow an unbiased exploration of the two phases. To
cope with this problem, it has recently been proposed in
Ref.~\cite{gul} to consider the whole set of experimental events
reweighting them by the excitation energy distribution. This
reweighting procedure produces a statistical ensemble which is
intermediate between the microcanonical and the canonical ensemble.
It is shown in Ref.~\cite{gul} that, within simple models, the
convexity of the order parameter distribution in this experimentally
accessible statistical ensemble can be associated to the convexity
of the underlying entropy.

\begin{figure}[hb]
\begin{center}
\resizebox{0.8\columnwidth}{!} {\includegraphics{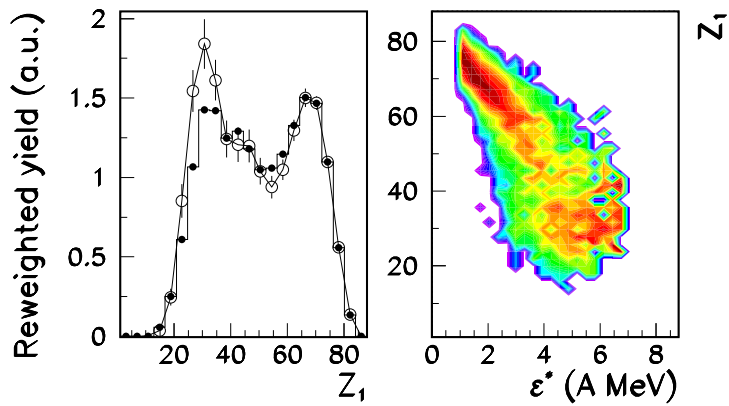}}
\caption{(Color online) \ \ \ 
Left panel: distribution of Z$_1$ for a reweighted
excitation energy distribution (open circles). The full points are
obtained in the case of constraint on the velocity of the largest
fragment. The lines are drawn to guide the eye. Right panel:
bi-dimensional plot of Z$_1$ vs. the excitation energy distribution
for the re-weighted distribution} \label{phase}
\end{center}
\end{figure}

This simple procedure allows to get rid of the trivial entrance
channel impact parameter geometry that naturally favors the lower
part of the excitation energy distribution. To produce a flat
excitation energy distribution we have therefore reweighted the
Z$_1$ yields in each $\varepsilon$* bin with a factor proportional
to the inverse of the bin statistics.
The results are given in Fig~\ref{phase}, which shows the
distribution of the largest cluster charge and a bi-dimensional plot
of the largest cluster charge {\it vs.} the excitation energy for
the reweighted distribution. A bimodal behavior of the largest
charge clearly emerges. As we have already mentioned, a first order
phase transition should imply a non zero latent heat, meaning that
the two "phases" observed at the same "temperature" should be
associated to different excitation energies.
We can see in the right panel of Fig.~\ref{phase} that indeed the
two decay modes observed in the $Z_1$ distribution correspond to
different values of the calorimetric excitation energy: the cut $Z_1
\approx$ 50 that roughly identifies the two bumps (see
Fig.~\ref{whole}) succeeds in splitting the energy distributions
into two separate components. The lower (higher) $Z_1$ component
roughly corresponds to a deposited energy higher (lower) than $\sim
3.5$ AMeV. Also shown in Fig.~\ref{phase} is the effect of the
condition on the velocity of the largest fragment on the bimodal
distribution. Clearly this condition cuts the events on the high
energy side, but does not affect at all the bimodal distribution.

If we interpret the two $Z_1$ bumps as two coexisting phases, it
would be tempting to estimate the latent heat of the transition from
the energy distance between the two peaks. The "liquid" peak points
to an excitation energy $\varepsilon_1$*$\approx 2$~AMeV which
nicely agrees with the indication of the fluctuation measurement
shown in the right part of the Fig.~\ref{cneg}. The "vapor"
contribution peaks at $\varepsilon_2$*$\approx 5.5$~AMeV, a slightly
lower value compared to the location of the second divergence in the
fluctuation analysis. This discrepancy may be due to the intrinsic
limitations of the reweighting procedure, that does not allow
sufficient energy fluctuations as compared to a physical heat
bath~\cite{gul}; it may also point to an incomplete exploration of
the high energy phase space in our data sample, that cuts the
distributions on the high energy side. From the temperature
estimations obtained for this data set~\cite{critical}, the "liquid"
peak is associated to T$_\ell$~=~4~$\pm$~0.5~MeV, while the vapor
contribution corresponds to T$_v$~=~4.7~$\pm$~0.5~MeV. This result
is consistent with the expectations from a first order phase
transition smoothed by finite size effects.

 To summarize, the results of Fig.~\ref{phase}
suggest that the observed sudden change from evaporation to
multifragmentation is the finite size precursor of a first order
phase transition. Higher statistics samples obtained with collisions
at higher beam energy could allow to be quantitatively more
conclusive about the compatibility between fluctuations and
bimodality~\cite{bonnet}. In addition a detailed study of the
convexity properties of the distributions would be very
welcome~\cite{gul,bonnet2}.

\section{Conclusions}

 In this paper we have presented a new analysis of the
35 AMeV quasi-projectile Au+Au data collected with the
Multics-Miniball apparatus, studying the distributions of the
largest cluster charge emitted in the de-excitation of a constant
size quasi-projectile source within a large range of dissipated
energy. An evident transition from an evaporative to a
multifragmentation pattern has been observed. Once the trivial
entrance channel effect of the impact parameter has been removed by
reweighting the Z$_1$ distribution by the statistics of the
excitation energy distribution, a bimodal behavior emerges. Such a
behavior supports the interpretation of this "discontinuity" of the
de-excitation mode as the finite system counterpart of a first order
phase transition.

As already pointed out, the bimodality signal {\it per se} cannot be
considered a clear signal of first order phase
transition~\cite{lop,aich}.   More data are needed in order to study
this signal also in central collisions~\cite{indra}.   Nevertheless our
data have shown a variety of different signals that are
coherently pointing to a first order liquid-gas-like phase transition.
We recall the determination of thermodynamically consistent critical
exponents, both in a moment analysis~\cite{dag} and in an analysis
"\`a la Fisher"~\cite{critical}, and the fluctuation peak in the
partial energy distribution, with a strength of fluctuations
consistent with the existence of a negative branch for the
microcanonical heat capacity~\cite{rely,negative}. 

Finally we would like to stress that, besides some small differences
due to the different energy range, the trend of our data are
consistent with the recent findings~\cite{indra,bonnet} for a system of
similar size at higher incident energies.
More quantitatively, the value of the higher $Z_1$ bump found in
Refs.\cite{indra,bonnet} is fully compatible with the one of this work.
Conversely, the lower $Z_1$ peak is found in Ref.\cite{indra}
at $Z_1\approx 15$, which is a lower value than the one presented
in Fig.\ref{phase}. This is most probably an effect of the decreasing
quasi-projectile average source size with increasing dissipation, with a non
negligible contribution of matter stopped close to midrapidity~\cite{indra}. 
Indeed if in this same Indra data-set a constant source size is explicitly imposed 
in the data selection~\cite{bonnet}
(similarly to our analysis, see Fig.\ref{cneg}), 
a higher value $Z_1\approx 25$ compatible with Fig.\ref{phase} is found.

%

\begin{thebibliography}{}
%
\bibitem{history} G.F. Bertsch and P.J. Siemens, Phys. Rev. Lett 126 (1983)
9.
\bibitem{SMM} J. P. Bondorf, A. S. Botvina, A. S. Iljinov, I. N. Mishustin, K. Sneppen, Phys. Rep. 257
(1995) 133.
\bibitem{gross}D.H.E. Gross, {\it Microcanonical Thermodynamics:
Phase Transitions in Small Systems}, Lecture Notes in Physics 66,
World Scientific, Singapore (2001).

\bibitem{rass} A. Bonasera, M. Bruno, C. Dorso, P.F. Mastinu, Riv.
Nuovo Cim. vol.23, n.2 (2000).

\bibitem{chomaz} Ph. Chomaz, {\it The Nuclear Liquid Gas Phase transition and Phase Coexistence},
Int. Nucl. Phys. Conference INPC 2001, AIP Proceedings Vol. No. 610,
2002.

\bibitem{rely} M. D'Agostino {\it et al.}, Nucl. Phys. A699 (2002)795.

\bibitem{det} I. Iori {\it et al.}, Nucl. Instr. and Meth A325
(1993) 458. ; R. T. DeSouza {\it et al.}, Nucl. Instr. and Meth A295
(1990) 109

\bibitem{praga} M. D'Agostino {\it et al.}, Nucl. Phys. A749 (2005) 55.

\bibitem{dag} M. D'Agostino {\it et al.}, Nucl. Phys. A650, 329
(1999); M. D'Agostino {\it et al.}, Proc. of {\it XXXVIII Int.
Winter Meeting on Nuclear Physics}, Bormio 2000, Ric. Sci. Univ.
Perm. 116 (2000) 386 ; Proc. of {\it Bologna 2000: Structure of the
Nucleus at the Dawn of the Century}, G.C. Bonsignori, M. Bruno, A.
Ventura, D. Vretenar (eds.), World Sci. 2001, 215.

\bibitem{mil} P. Milazzo {\it et al.}, Phys. Rev. C 58, 953 (1998).

\bibitem{caloric} J. Pochodzalla {\it et al.}, Phys. Rev. Lett. 75, 1040 (1995).

\bibitem{critical} M. D'Agostino {\it et al.}, Nucl. Phys. A724 (2003) 455.

\bibitem{negative} M. D'Agostino {\it et al.}, Phys. Lett. B473,
219(2000).
\bibitem{negth} Ph. Chomaz, F. Gulminelli, Nucl. Phys. A647 (1999)
153.

\bibitem{ma} Y. G. Ma {\it et al.}, Phys. Rev. C71 (2005) 054606.
\bibitem{coll}  J. B. Natowitz {\it et al.}, Phys. Rev. C 65, 034618
(2002); J. B. Elliott {\it et al.}, Phys. Rev. Lett. {\bf 85}, 1194
(2000); J. B. Elliott {\it et al.}, Phys. Rev. C 67 (2003) 024609.

\bibitem{dag2} M. D'Agostino {\it et al.}, Nucl. Phys. A734 (2004)
512.

\bibitem{bruno} M. Bruno {\it et al.} Proc. Fifth Italy-Japan Symposium, Naples, November 2004
in G. La Rana, C. Signorini, S. Shimura (Eds.), {\it Recent
Achievements and Perspectives in Nuclear Physics}, W. Sci. (2005)
209.

\bibitem{leneindre} N. Leneindre {\it et al.}, Nucl. Phys. A795
(2007) 47.

\bibitem{negclu} M. Schmidt {\it et al.} Phys. Rev. Lett. {\bf 79}, 99 (1997);
M. Ery\"urek and M.H. G\"uven, Physica A 337 (2007) 514.
\bibitem{neg3} F. Gobet {\it et al.} Phys. Rev. Lett. 89, 183403 (2002).

\bibitem{PRE} Ph. Chomaz, F. Gulminelli, V. Duflot, Phys. Rev. E
64 (2001) 046114; F. Gulminelli, Ann. de Phys. 29 (2004) 1;

\bibitem{binder} K. Binder and D. P. Landau, Phys. Rev. B 30, 1477 (1984).

\bibitem{dasgupta} G. Chaudhuri, S. Das Gupta, Phys. Rev. C 76
(2007) 014619.

\bibitem{zeroes} S. Grossmann and W. Rosenhauer, Z. Phys. 207
(1967) 138; P. Borrmann {\it et al.}, Phys. Rev. Lett. 84 (2000)
3511; K. C. Lee, Phys. Rev. E 53 (1996) 6558; Ph. Chomaz and F.
Gulminelli, Physica A 330 (2003) 451.

\bibitem{intruder} M.S. Challa and J. H. Hetherington, Phys. Rev.
Lett. 60 (1988) 77; Phys. Rev. A 38 (1988) 6324.

\bibitem{WCI} O. Lopez and M.F. Rivet, in {\it Dynamics and Thermodynamics with Nuclear Degrees of Freedom}, Eur. Phys. J. A 30 (2006)
263.

\bibitem{indra} M. Pichon {\it et al.}, Nucl. Phys. A 779 (2006)
267.

\bibitem{bell} N. Bellaize {\it et al.}, Nucl. Phys. A 709 (2002) 367.
\bibitem{traut} W. Trautmann, Proc. of {\it XLIII Int. Winter Meeting on Nuclear Physics},
Bormio 2007.

\bibitem{campi} X. Campi, J. Phys A 19 (1986) 917; R. Botet, M. Ploszajczak, Phys. Rev. E 62 (2000) 1825.

\bibitem{big} F. Gulminelli and Ph. Chomaz, Phys. Rev. C 71 (2005)
054607.
\bibitem{lop} O. Lopez {\it et al.}, Phys. Rev. Lett.95 (2005)
242701.

\bibitem{aich}A. Le Fevre, J. Aichelin, Phys. Rev. Lett 100 (2008) 042701.

\bibitem{cugn} J. Cugnon, D. L'Hote, Nucl. Phys. A 397 (1983)
519.
\bibitem{lecolley} J. F. Lecolley {\it et al.}, Nucl. Instr. and
Meth. A 441 (2000) 517.

\bibitem{win} R.A. Winyard {\it et al.}, Nucl. Instr. and Meth. 95 (1971) 141.

\bibitem{bonnet} E. Bonnet, PhD thesis, Universite' Paris VI, 2006,
http://tel.archives-ouvertes.fr/tel-00121736.

\bibitem{rivet} M.~F.~Rivet~{\it et al.},~nucl-ex/0205010~IWM2001~{\it Int. Work. on Multifragmentation and Related Topics}, Catania, nov.
2001; B. Borderie, J. Phys. G : Nucl. Part. Phys. 28 (2002) R217.

\bibitem{gul} F. Gulminelli, Nucl. Phys. A791 (2007) 165.

\bibitem{bonnet2} E.Bonnet, B.Borderie et al., in preparation.


\end{thebibliography}
%

\end{document}